\documentclass[aps,twocolumn,showpacs,preprintnumbers,amsmath,amssymb]{revtex4}
 \usepackage{graphicx}

\begin{document}
\title{$\bar{K}\bar{K}N$ molecule
state in three-body calculation}

\author{Yoshiko Kanada-En'yo and Daisuke Jido}
\address{Yukawa Institute for Theoretical Physics, Kyoto University,
Kyoto 606-8502, Japan}

\begin{abstract}
A $\bar{K}\bar{K}N$ system with $I=1/2$ and $J^P=1/2^+$ 
is investigated with 
non-relativistic 
three-body calculations by using effective $\bar{K}N$ and $\bar K \bar K$ 
interactions. The present investigation suggests that 
a weakly bound state for the $\bar{K}\bar{K}N$ system can be formed 
below the two-body threshold of $\bar{K}$ and quasibound $\bar{K}N$
with a $40 \sim 60$ MeV decay width of 
$\bar{K}\bar{K}N\rightarrow \pi Y\bar{K}$. This corresponds to
an excited $\Xi$ baryon with $J^P=1/2^+$ located around 1.9 GeV.
Studying the wave function of the $\bar{K}\bar{K}N$ system obtained 
in this formulation, we find that 
the three-body bound system has a characteristic structure 
of $\bar K N$+$\bar K$ cluster with spatial extent.
\end{abstract}
\maketitle

\noindent

\section{Introduction}

The study of hadron structure is one of the most important issues
in hadron physics. Recent interest in this line is developed in exploring 
quasi-bound systems of mesons and baryons governed by strong 
interactions among the hadrons. One of the long-standing candidates is the 
$\Lambda(1405)$ resonance considered as a quasi-bound state of $\bar KN$ 
system~\cite{Dalitz:1959dn}. It has been also suggested that 
the $f_{0}(980)$ scalar meson is a molecular state
of $\bar KK$~\cite{Weinstein:1982gc}. 
For nuclear systems, bound states of an eta meson in nuclei and an anti-kaon 
in light nuclei were predicted in Refs.~\cite{Haider:1986sa,Kishimoto:1999yj,akaishi02}.
Recently a multi-hadron state was proposed for  the $\Theta^{+}$ baryon resonance 
to explain its narrow width~\cite{Kishimoto:2003xy}. 

In such multi-hadron systems, anti-kaon plays a unique and important
role due to its heavy mass and Nambu-Goldstone boson nature. 
The heavier kaon mass indicates
stronger $s$-wave interactions around the threshold
than those for pion according to the chiral effective theory. 
In addition, since typical kaon kinetic energy in the system estimated 
by range of hadronic interaction is small in
comparison with the kaon mass,
we may treat kaons in multi-hadron systems in many-body formulations.

The strong attraction in the $\bar K N$ system led to the idea
of deeply bound kaonic states in light nuclei, such as $K^-pp$ and 
$K^-ppn$, by Akaishi and Yamazaki~\cite{akaishi02,yamazaki02,Akaishi:2005sn,Yamazaki:2003hs,Dote:2003ac}.
Later, many theoretical studies on the structure of the $K^-pp$ system 
have been done, for example in
Faddeev calculations~\cite{shevchenko07,ikeda07} and in variational 
calculations~\cite{yamazaki07,dote08}, having turned out that the $K^-pp$
system is bound with a large width.
Experimental search for these states has been reported~\cite{Agnello:2005qj,Suzuki:2004ep,Sato:2007sb,Suzuki:2007kn}, 
whereas clear evidences are not obtained yet and
interpretations of the experimental data are
controversial~\cite{Magas:2006fn,Yamazaki:2006yc}.
Motivated by the strong 
$\bar{K}N(I=0)$ attractions, quests of multi-$\bar{K}$ nuclei are 
challenging issues~\cite{Yamazaki:2003hs,Dote:2005nb,Gazda:2007wd,Gazda:2008xd,Muto:2007zz}. 
This is also one of the key subjects related to
kaon condensation in dense nuclear matter 
\cite{Kaplan86,Muto:1992gb,Muto:1992gc}.
%

The key issue for the study of the $\bar KN$ interaction is the resonance position of the $\Lambda(1405)$ in the $\bar KN$ scattering amplitude. The $\Lambda(1405)$ is observed around 1405 MeV in the $\pi \Sigma$ final state interaction, as summarized in the particle data group~\cite{PDG}. Based on this fact, a phenomenological effective
$\bar K N$ potential (AY potential) was derived in Refs.~\cite{akaishi02,yamazaki07},
having relatively strong attraction in the $I=0$ channel to provide the $K^-p$ bound state at 1405 MeV. 
Recent theoretical studies of the $\Lambda(1405)$ based on chiral unitary approach
have indicated that 
the $\Lambda(1405)$ is described as a superposition of two pole states
and one of the state is considered to be a $\bar{K}N$ quasibound state 
embedded in the strongly interacting $\pi\Sigma$ continuum~\cite{Oller:2000fj,Jido:2002yz,Jido:2003cb,Hyodo:2007jk,Hyodo:2008xr,Fink:1989uk}.
This double-pole conjecture suggests  that the resonance position in the
$\bar{K}N$ scattering amplitudes with $I=0$ is around 1420 MeV, 
which is higher than the energy position of the 
nominal $\Lambda(1405)$ resonance. 
Based on this chiral SU(3) coupled-channel dynamics,
Hyodo and Weise have derived another effective $\bar{K}N$ 
potential (HW potential) \cite{Hyodo:2007jq}.
The HW potential provides a $\bar{K}N$ quasibound state at 
$\sim$ 1420 MeV instead of 1405 MeV,
and it is not as strong as the AY potential.

The common achievement of these studies 
on the $s$-wave $\bar K N$ effective potential
is that the $\bar{K}N$ interaction with $I=0$ is strongly attractive 
and describes the $\Lambda(1405)$ resonance as 
a $\bar{K}N$ quasibound state. 
Basing on these strong $\bar KN$ interactions, we 
examine possible bound states of a lightest two-anti-kaon nuclear system, namely
$\bar{K}\bar{K}N$ with $I=1/2$ and $J^P=1/2^+$, in the hadronic molecule picture. 
We expect that the three-body $\bar K \bar K N$ system will form a bound state
due to the strong attraction in the two-body $\bar KN$ subsystem producing the
$10\sim 30$ MeV binding energy. The questions raised here are 
whether the strong attraction is 
enough for a three-body bound state below the threshold of the $\bar K$ and
the $\bar KN$ quasibound state, and what structure the bound state has, 
if it is formed.

The $\bar K \bar K N$ molecule state may have characteristic 
decay patterns depending on the binding energy. 
If the three-body state is above the threshold of
the $\bar{K}$ and the $\bar{K}N$ quasibound state, 
it can decay into $\bar K$ and $\Lambda(1405)$, and, consequently, 
the bound state  has a large width. 
If the $\bar K \bar K N$ system is bound below the threshold, 
the bound state  has  a comparable width 
to the $\Lambda(1405)$, and the main decay mode is a three-body 
$\pi \Sigma \bar K$ and two-hadron decays are strongly suppressed 
in contrast to usual excited baryons. 
For deeply bound $\bar K \bar K N$ systems, the molecular picture may be 
broken down and two-hadron decays may be favored. 
%
%
Presently experimental data for $S=-2$ channel are very poor. But we expect that  
such an interesting molecule state could be observed as an exited state of $\Xi$ baryon,
for example, in double strangeness exchange $(K^-,K^+)$ processes at J-PARC.

In this paper, we investigate 
the $\bar{K}\bar{K}N$ system with $I=1/2$ and $J^P=1/2^+$ 
with non-relativistic three-body calculations 
by using the HW and AY potential as effective $\bar{K}N$ interactions.
In Sec.~\ref{sec:formulation}, we describe the framework 
of the present calculations. We apply a variational approach with 
a Gaussian expansion method~\cite{Hiyama03} to 
solve the Schr\"odinger equation of the three-body system.
By treating the imaginary potentials perturbatively,  
we find the $\bar{K}\bar{K}N$ quasibound state. 
In Sec.~\ref{sec:results}, we present our results of the three-body
calculation. In analysis of the wave functions,
we discuss the structure and the binding mechanism of the 
$\bar{K}\bar{K}N$ state.
The effects of $\bar{K}\bar{K}$ interactions
on the $\bar{K}\bar{K}N$ system are also discussed.
Section~\ref{sec:summary} is devoted to summary and concluding remarks. 

\section{Framework} \label{sec:formulation}
We consider a non-relativistic three-body potential model for 
the $\bar{K}\bar{K}N$ system. 
We calculate the $\bar{K}\bar{K}N$ wave function 
by using effective two-body interactions~\cite{yamazaki07,Hyodo:2007jq} 
in local potential forms.
We apply a variational approach with a 
Gaussian expansion method~\cite{Hiyama03} in solving the Schr\"odinger equation
for the three-body system.
In this section we explain the details of the 
framework and interactions used in this work.

\subsection{Hamiltonian}

\begin{figure}[t]
\centerline{\includegraphics[width=7.5cm]{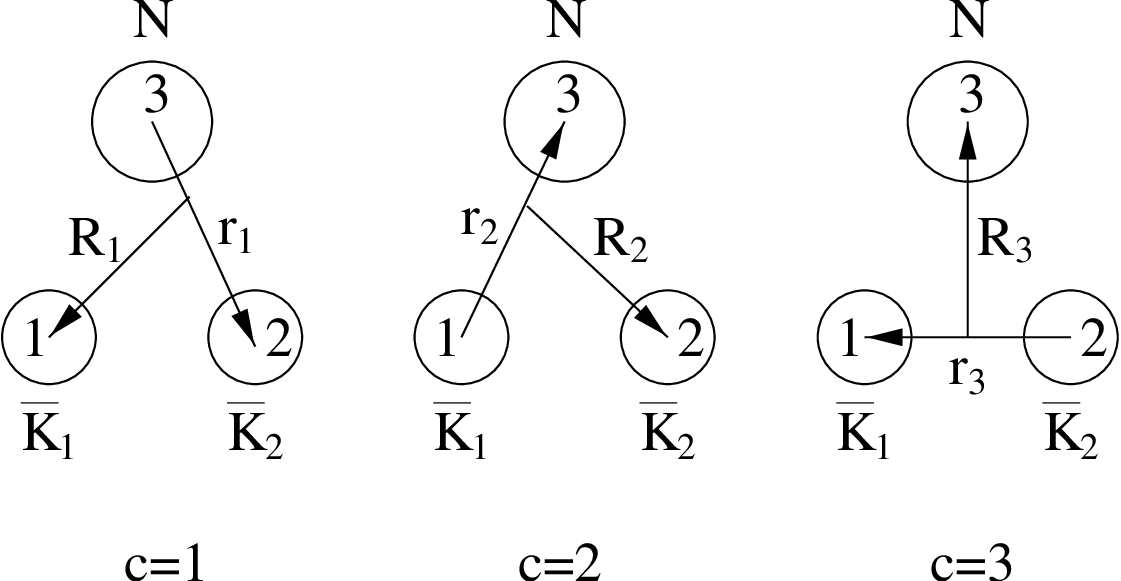}}
\caption{\label{fig:jacobi}
Three Jacobian coordinates of the $\bar{K}\bar{K}N$ system.
}
\end{figure}

In the present formulation, we use the Hamiltonian for 
the $\bar{K}\bar{K}N$ system given by
\begin{equation}\label{eq:hamiltonian}
H=T+V_{\bar{K}N}(r_1)+V_{\bar{K}N}(r_2)+V_{\bar{K}\bar{K}}(r_3),
\end{equation} 
with the kinetic energy $T$, the effective $\bar{K}N$ interaction
$V_{\bar{K}N}$ and  the $\bar{K}\bar{K}$ interaction $V_{\bar{K}\bar{K}}$.
These interactions are given by local potentials as functions of 
$\bar{K}$-$N$ distances, $r_{1}$, $r_{2}$, and the $\bar{K}$-$\bar{K}$ 
distance, 
$r_{3}$, which are defined by  
$r_{1}=|{\bf x}_2-{\bf x}_3|$, $r_{2}=|{\bf x}_3-{\bf x}_1|$ and   
$r_{3}=|{\bf x}_2-{\bf x}_1|$. 
The vectors ${\bf x}_1$, ${\bf x}_2$, ${\bf x}_3$ denote
the spatial coordinates of the first anti-kaon ($\bar{K}_1$),
the second anti-kaon ($\bar{K}_2$) and the nucleon, respectively. For convenience,
we use Jacobian coordinates,
${\bf r}_c$ and ${\bf R}_c$, in three rearrangement 
channels $c=1,2,3$ as shown in Fig.~\ref{fig:jacobi}.
In the present calculation, 
we assume isospin symmetry in the effective interactions, and we also
neglect the mass differences between $K^{-}$ and $\bar K^{0}$ and 
between proton and neutron by
using the averaged masses. 
We do not consider three-body forces nor transitions to two-hadron decays. 

The kinetic energy $T$ is simply expressed by the Jacobian coordinates 
for a rearrangement channel as 
\begin{equation}
T\equiv \frac{-1}{2\mu_{r_c}}\nabla_{r_c}^2
+\frac{-1}{2\mu_{R_c}}\nabla_{R_c}^2,
\end{equation} 
with the reduced masses $\mu_{r_c}$ and $\mu_{R_c}$ 
for the corresponding configuration.
For instance, 
$\mu_{r_{1}}=M_K M_N/(M_K+M_N)$ and 
$\mu_{R_{1}}=M_K (M_K+M_N)/(2M_K+M_N)$ for the rearrangement channel 
$c=1$.
Here $M_{N}$ and $M_{K}$ denote the averaged nucleon and kaon masses,
respectively, as $M_N=938.9$ MeV and $M_K=495.7$ MeV.

For the effective interactions, $V_{\bar KN}$ and $V_{\bar K \bar K}$, 
we use local potentials obtained by 
$s$-wave two-body scattering amplitudes with isospin symmetry.
In the present three-body calculations, 
we take $l$-independent potentials for simplicity.
The details of the local potentials will be discussed in Sec.~\ref{sec:effint}.
The coupled-channel effects of the $\bar KN$ to other relevant channels 
have been already implemented to the effective single-channel $\bar KN$ interaction 
$V_{\bar KN}$. Consequently $V_{\bar{K}N}$ has an imaginary part  
owing to scattering states below the $\bar KN$ threshold, and
the present Hamiltonian is not hermitian.
In the calculations of $\bar{K}\bar{K}N$ wave functions,
we first use only the real part of $V_{\bar{K}N}$ in a variational approach,
and then we calculate the energy $E$ with the expectation value 
of the total Hamiltonian~(\ref{eq:hamiltonian}) with respect to the obtained wave functions.
The width of the bound state is evaluated by the imaginary part of the energy as
$\Gamma=-2{\rm Im}E$.
The details of the calculational procedure will be described later.

\subsection{Effective interactions}
\label{sec:effint}
In this subsection, we explain the details of the effective interactions of the $\bar KN$ and $\bar K \bar K$ two-body subsystems in our formulation. 
We use two effective  $\bar KN$ 
interactions which were derived in different ways,
and compare the results obtained by these interactions to estimate 
theoretical uncertainties. Both $\bar KN$ interactions have so strong 
attraction as to provide the $\Lambda(1405)$
as a quasibound state of the $\bar KN$ system. The important difference
is the binding energy of the $\bar K N$ system, as already mentioned in introduction. 

One of the $\bar KN$ interactions which we use here is given by Hyodo and Weise in Ref.~\cite{Hyodo:2007jq}. 
This effective interaction was derived  based on 
the chiral unitary approach~\cite{ChUA} for $s$-wave scattering amplitude with strangeness $S=-1$ of the lowest-lying octet meson and baryon.
A equivalent single-channel $\bar KN$ amplitude can be obtained by 
reducing the four (five)-channel problem for $I=0$ ($I=1$) to a
$\bar KN$ single-channel problem including dynamics of the rest channels.  
The local $\bar KN$ potential was constructed in coordinate space so as 
to reproduce the single-channel $\bar KN$ interaction 
as a solution of the  Schr\"odinger equation for the $\bar KN$ system
with the local potential.

The potential is written in a one-range Gaussian form as
\begin{eqnarray}
V_{\bar{K}N} &=& U_{\bar{K}N}^{I=0}\exp\left[ -(r/b)^2\right]P_{\bar{K}N}(I=0)
\nonumber \\ && 
+U_{\bar{K}N}^{I=1}\exp\left[ -(r/b)^2\right]P_{\bar{K}N}(I=1),
\end{eqnarray}
with the isospin projection operator $P_{\bar{K}N}(I=0,1)$ and
the range parameter $b=0.47$ fm. 
The range parameter was optimized for  the parametrization referred as HNJH~\cite{Hyodo03} in Ref.~\cite{Hyodo:2007jq}.
We refer this potential as ``HW-HNJH potential''.
The strength $U_{\bar{K}N}(I=0,1)$ has energy dependence and is parametrized
in terms of a third order polynomial in the energy $\omega$: 
\begin{eqnarray}
\lefteqn{ U_{\bar{K}N}^{I=0,1}(\omega)=} && \nonumber \\
&& K^{I=0,1}_0+K^{I=0,1}_1 \omega+K^{I=0,1}_2 \omega^2
+K^{I=0,1}_3 \omega^3,
\end{eqnarray}
for 1300 MeV $\le \omega \le$ 1450 MeV.
The coefficients $K^{I=0,1}_i$ for $I=0$ and $I=1$ are given in Table IV and V 
of Ref.~\cite{Hyodo:2007jq}, respectively. Here we use
the corrected version of the local potentials~\cite{Hyodo:2007jq}.

In chiral unitary approaches for the meson-baryon interactions, 
two poles are generated at 1.4 GeV region, and 
the $\Lambda(1405)$ resonance is described 
as a $\bar{K}N$ quasibound state in the strongly interacting
$\pi\Sigma$ continuum~\cite{Oller:2000fj,Jido:2003cb}. 
In this case, the peak position of the
$\bar{K}N$ scattering amplitudes in the $I=0$ channel appears at 
$\omega\sim 1420$ MeV, which is higher than the energy position of the 
$\Lambda(1405)$ resonance observed in the final state interaction of the 
$\pi\Sigma$ channel.
The HW potential reproduces this feature of $\bar{K}N$ scattering amplitudes,
and therefore the position of the $\bar{K}N$ quasibound state is located at 
$\omega\sim 1420$ MeV.
With the HW-HNJH potential,
$\Lambda(1405)$ can be calculated 
as a $\bar{K}N$ quasibound state
by solving the $\bar{K}N$ two-body Schr\"odinger 
equation in the $s$-wave $I=0$ channel. 
We get a $\bar{K}N$ quasibound state at 1423 MeV with a
width $\Gamma=44$ MeV,  treating the imaginary part of the potentials perturbatively.
This is a weakly bound $\bar{K}N$ state located at 
11 MeV below the $\bar{K}N$ threshold ($M_K+M_N=1434$ MeV). 
The root-mean-square of the 
$\bar{K}$-$N$ distance ($d_{\bar KN}$) is also calculated as 1.9 fm.

The HW-HNJH potential $V_{\bar{K}N}^{I=0,1}(\omega)$
depends on the energy of the subsystem $\bar{K}N$.
We regard, however, $\omega$ as
an interaction parameter and use fixed values of 
$\omega$,
since the energy dependence is small in the region of interest, 
$\omega=1400 \sim 1440$ MeV.
We take two values of 
$\omega=M_N+M_K-\delta\omega$ with $\delta\omega=$0 and 11 MeV.
The choice of $\delta\omega=0$ MeV corresponds to  $\bar{K}N$ at the threshold,
while $\delta\omega=11$ MeV is for 
the binding energy $B(\bar{K}N)= 11$ MeV 
of the $\bar{K}N$ bound state obtained by the 
HW-HNJH potential.
In each of the fixed HW potential,
we calculate the energy $E^{\rm Re}=-B(\bar{K}\bar{K}N)$ 
of the $\bar{K}\bar{K}N$ system and also the 
energy $-B(\bar{K}N)$ of the $\bar{K}N$ system.

The other $\bar KN$ interaction which we use is given by Akaishi and Yamazaki (AY)
in Refs.~\cite{akaishi02,yamazaki07}. The AY potential was derived in a phenomenological way to start with the ansatz that the 
$\Lambda(1405)$ resonance is a $K^-p$ bound state at 1405 MeV as reported by PDG~\cite{PDG}.
The AY potential is energy independent and was parametrized so as to reproduce
the $\bar KN$ quasibound state at the PDG values of the $\Lambda(1405)$:
\begin{eqnarray} 
&V_{\bar{K}N}^{I=0}(r)=(-595-83i)\exp \left[-(r/0.66 {\rm fm})^2 \right],\\
&V_{\bar{K}N}^{I=1}(r)=(-175-105i)\exp \left[-(r/0.66 {\rm fm})^2 \right].
\end{eqnarray} 
The AY potential has a stronger attraction in the $I=0$ channel
because it was adjusted to generate a $\bar{K}N(I=0)$ 
quasibound state at $\sim$ 1405 MeV. It corresponds to 
the $\bar{K}N$ binding energy $B(\bar{K}N)\sim 30$ MeV which is
much larger than $B(\bar{K}N)=11$ MeV for the HW-HNJH potential.
By solving the $\bar{K}N$ two-body Schr\"odinger 
with a perturbative treatment of the imaginary potential, 
we get a $\bar{K}N$ $(I=0)$ state at 1403 MeV with a
width of 40 MeV, and 
the root-mean-square $\bar{K}$-$N$ distance is $d_{\bar KN}=1.4$ fm.
Because of the strong attraction in the $I=0$ channel of the AY potential, 
the $\bar{K}N(I=0)$ quasibound state has the deeper binding and 
smaller radius than those given by the HW-HNJH potential.
We comment that the size of the $\bar K$-$N$ state 
is sensitive to the binding energy~\cite{Sekihara:2008qk}.

For the $\bar{K}\bar{K}$ interactions, there are few experimental data.
The low energy theorem based on the current algebra suggests a repulsive
interaction. For the $s$-wave interaction of $\bar K \bar K$, the $I=0$ is 
forbidden due to Einstein-Bose statistics.
Thus we assume $V^{I=0}_{\bar{K}\bar{K}}=0$.
We introduce the effective interaction of the subsystem $\bar K \bar K$ with $I=1$, $V^{I=1}_{\bar{K}\bar{K}}$,  in a Gaussian form
\begin{equation}
V^{I=1}_{\bar{K}\bar{K}}(r)=U^{I=1}_{\bar{K}\bar{K}}
\exp\left[ -(r/b)^2\right]P_{\bar{K}\bar{K}}(I=1),
\end{equation}
where the range parameter $b$ is chosen to be the same value as that of the 
$\bar{K}N$ interaction. The interaction $V^{I=1}_{\bar{K}\bar{K}}$ is a real
function, since there are no decay channels open for the $\bar K \bar K$ system. 

The strength $U^{I=1}_{\bar{K}\bar{K}}$ is estimated by theoretical calculations
of the scattering length of $K^{+}K^{+}$, which is equivalent to that of $\bar K \bar K$ with $I=1$.
Recently, the $K^+K^+$ scattering length has been obtained in 
lattice QCD calculation as
$a_{K^+K^+}=-0.141\pm 0.006$~fm~\cite{Beane07}. 
This value is consistent with the leading order calculation of the chiral perturbation
theory, $a_{K^{+}K^{+}}=-0.147$ fm, which is obtained by
\begin{equation}
   a_{K^{+}K^{+}} = -\frac{m_{K}}{16 \pi f_{K}^{2}} ,
\end{equation}
with $f_{K}=115$ MeV. In the present calculation, 
the strength $U^{I=1}_{\bar{K}\bar{K}}$ is adjusted to reproduce
the scattering length $a^{I=1}_{KK}=-0.14$ fm.
For the HW-HNJH potential, with the interaction range $b=0.47$ fm, we obtain
$U^{I=1}_{\bar{K}\bar{K}}=313$ MeV, and we find
$U^{I=1}_{\bar{K}\bar{K}}=104$ MeV with $b=0.66$ fm 
for the AY potential. We will also try a weaker repulsion
with $a^{I=1}_{KK}=-0.10$ 
using $U^{I=1}_{\bar{K}\bar{K}}=205$ MeV and 70 MeV
for the HW-HNJH and the AY potentials, respectively.
In order to examine how strong the $\bar KN$ interaction 
for the three body system,
at first we neglect the $\bar K \bar K$ repulsion, 
then we see the effect of the repulsive 
interaction with including the effective $\bar K \bar K$ 
interaction given above. 
 
\subsection{Three-body wave function}\label{subsec:wf}

The three-body $\bar{K}\bar{K}N$ wave function $\Psi$
is described 
as a linear combination of amplitudes 
$\Phi^{(c)}_{I_{KK}}({\bf r}_{c}, {\bf R}_{c})$
of two rearrangement channels 
$c=1,3$ (Fig.~\ref{fig:jacobi}), since
the channel $c=2$ is included 
by symmetrization of the two anti-kaon in the wave function.  
In the present calculation, we take the model space limited to
$l_c=0$ and $L_c=0$ of the orbital-angular momenta for the
Jacobian coordinates ${\bf r}_c$ and ${\bf R}_c$ in the 
channel $c$. 
Then the wave function of the $\bar{K}\bar{K}N$ system 
with $I=1/2$ and $J^P=1/2^{+}$ is written as
\begin{eqnarray}
 \Psi &=& \frac{1+P_{12}}{\sqrt{2}}\Phi,\\
\Phi&=&
\Phi^{(c=1)}_{I_{KK}=0}({\bf r}_1, {\bf R}_1) 
\left[[\bar{K}\bar{K}]_{I_{KK}=0}N\right]_{I=1/2} \nonumber \\
&+& \Phi^{(c=1)}_{I_{KK}=1}({\bf r}_1, {\bf R}_1) 
\left[[\bar{K}\bar{K}]_{I_{KK}=1}N\right]_{I=1/2}\nonumber \\
&+& \Phi^{(c=3)}_{I_{KK}=1}({\bf r}_3, {\bf R}_3) 
\left[[\bar{K}\bar{K}]_{I_{KK}=1}N\right]_{I=1/2}, \label{eq:wavefunc}
\end{eqnarray}
where $P_{12}$ is the exchange operator between the two anti-kaons, $\bar K_1$ 
and $\bar K_2$ for two bosons.
The $\left[[\bar{K}\bar{K}]_{I_{KK}}N\right]_{I=1/2}$ specifies 
the isospin configuration of the wave function 
$\Phi^{(c)}_{I_{KK}}({\bf r}_c, {\bf R}_c)$, meaning that the total isospin $1/2$ for the $\bar K \bar K N$ system is given by combination of total isospin $I_{KK}$ for the $\bar K \bar K$ subsystem and isospin $1/2$ for the nucleon. The isospin configuration 
$\left[[\bar{K}\bar{K}]_{I_{KK}=0}N\right]_{I=1/2}$@in the $c=3$ is not
necessary, because it vanishes after the symmetrization in the case of 
$l_{3}=0$.

As mentioned above, we omit basis wave functions with 
$l_c\ge 1$ and $L_c\ge 1$ in each rearrangement channel. 
This is consistent with the fact that 
the effective local potentials used in the
present calculations are derived to reproduce the $s$-wave
two-body dynamics. We comment that components with 
non-zero angular momenta of two-body subsystems are 
contained in the model wave function
through rearrangement of three-body configurations, 
although the $s$-wave component
is expected to be dominant.

In solving the Schr\"odinger equation for 
the $\bar{K}\bar{K}N$ system,
we adopt the Gaussian expansion method for three-body systems given in
Ref.~\cite{Hiyama03}.
The spatial wave functions $\Phi^{(c)}_{I_{KK}}({\bf r}_c, {\bf R}_c)$
of the subcomponent of Eq.~(\ref{eq:wavefunc})
are expanded in terms of the Gaussian basis functions, $\phi^G_{n}({\bf r})$ and
$\psi^G_{n}({\bf R})$:
\begin{equation}
\Phi^{(c)}_{I_{KK}}({\bf r}_c, {\bf R}_c)=\sum_{n_c,N_c}^{n_{\rm max}, N_{\rm max}}
A^{c,I_{KK}}_{n_c,N_c} \phi^G_{n_c}({\bf r}_c)
\psi^G_{N_c}({\bf R}_c). \label{eq:gauss}
\end{equation}
The coefficients $A^{c,I_{KK}}_{n_c,N_c}$ are determined
by variational principle when we solve the Schr\"odinger equation. 
In Eq.~(\ref{eq:gauss}), $n_{\rm max}$ and $N_{\rm max}$ are 
the numbers of the Gaussian basis, and the basis functions
are defined by
\begin{eqnarray}
\phi^G_{n}({\bf r}) &=& {\cal N}_n e^{-\nu_n r^2},  \\
\psi^G_{n}({\bf R}) &=& {\cal N}_N e^{-\lambda_N R^2},
\end{eqnarray}
where the normalization constants are given by ${\cal N}_n=2(2\nu_n)^{3/4}\pi^{-1/4}$ 
and ${\cal N}_N=2(2\lambda_N)^{3/4}\pi^{-1/4}$, and 
the Gaussian ranges, $\nu_n$ and $\lambda_{N}$, are given by 
\begin{align}
&\nu_n=1/r_n^2,  & & r_n=r_{\rm min} \left( \frac{r_{\rm max}}{r_{\rm min}}
 \right)^{\frac{n-1}{n_{\rm max}-1}},\\
&\lambda_N=1/R_N^2, & & R_N=R_{\rm min} \left( \frac{R_{\rm max}}{R_{\rm min}}
 \right)^{\frac{N-1}{N_{\rm max}-1}}.
\end{align}
We take enough bases for the present system
by using the values given in Table~ \ref{tab:range}
for the basis numbers and range parameters in the channel $c=1$ 
and $c=3$.
We find that
the mixing effect of the rearrangement channel $c=3$ is 
very small in the present results of the $\bar{K}\bar{K}N$ system.
This is because the $\bar{K}\bar{K}$ interactions are not attractive 
and therefore the $\bar{K}$-$\bar{K}$ correlation is not strong 
in the $\bar{K}\bar{K}N$ system. This is different from the case of 
the $K^-pp$ system where the $p$-$p$ correlation is significant 
because of the attractive $NN$ interaction.

\begin{table}[ht]
\caption{\protect\label{tab:range} 
The numbers and range parameters of the basis functions 
for the rearrangement channels, $c=1$ and $c=3$. 
}
\begin{tabular}{rcccccc}
\hline
channel & $n_{\rm max}$ & $r_{\rm min}$ & $r_{\rm max}$ &
 $N_{\rm max}$ & $R_{\rm min}$ & $R_{\rm max}$ \\
  &  & (fm)  & (fm) &  &  (fm) & (fm) \\
\hline
$c=1$ & 15 & 0.2 & 20 & 25 & 0.2 & 200 \\ 
$c=3$ & 15 & 0.2 & 20 & 15 & 0.2 & 20 \\ 
\hline
\end{tabular}
\end{table}

\subsection{Procedure of calculations}
The wave function of the $\bar{K}\bar{K}N$ system is obtained by solving  
the Schr\"odinger equation:
\begin{equation}
\left[ T+V_{\bar{K}N}(r_1)+V_{\bar{K}N}(r_2)
+V_{\bar{K}\bar{K}}(r_3)-E\right]\Psi=0.
\end{equation}
The effective interaction $V_{\bar{K}N}$ is complex due to the presence of the
decay channels below the threshold, while $V_{\bar{K}\bar{K}}$ is expressed by
real numbers.

In order to solve this equation with variational principle,
we treat the imaginary part of the potentials perturbatively.
Separating the real part of the Hamiltonian, we write
\begin{eqnarray}
H^{\rm Re}= T+{\rm Re}V_{\bar{K}N}(r_1)+
{\rm Re}V_{\bar{K}N}(r_2)
+V_{\bar{K}\bar{K}}(r_3). 
\end{eqnarray}
We first calculate the wave function for the real part of the Hamiltonian, $H^{\rm Re}$,
with variational principle
in the model space of the Gaussian expansion described in 
Sec.~\ref{subsec:wf}. This is equivalent to determine the eigenenergy 
$E^{\rm Re}$ and the coefficients 
$A^{c,I_{KK}}_{n_c,N_c}$ of the Gaussian wave functions
$\phi^G_{n_c}({\bf r}_c)
\psi^G_{N_c}({\bf r}_c)
\left[[\bar{K}\bar{K}]_{I_{KK}}N\right]_{I=1/2}$
by diagonalizing the norm matrix and Hamiltonian matrix
\begin{widetext}
\begin{equation}
\left\langle \phi^G_{n'_c}({\bf r}_c)
\psi^G_{N'_c}({\bf r}_c)
\left[[\bar{K}\bar{K}]_{I'_{KK}}N\right]_{I=1/2} 
\left| H^{\rm Re} \right|
\phi^G_{n_c}({\bf r}_c)
\psi^G_{N_c}({\bf r}_c)
\left[[\bar{K}\bar{K}]_{I_{KK}}N\right]_{I=1/2} \right\rangle.
\end{equation}
\end{widetext}
After this variational calculation, we take 
the lowest-energy solution for $H^{\rm Re}$. 
The binding energy $B(\bar{K}\bar{K}N)$ of the three-body
system is given as $B(\bar{K}\bar{K}N)=-E^{\rm Re}$.
It should be checked if the bound state of the $\bar{K}\bar{K}N$
is lower than the threshold of the $\bar K$ and the quasibound state of 
$\bar KN$, because the $\bar{K}\bar{K}N$ solution obtained above
the $\bar{K}N+\bar K$ threshold
is not a three-body bound state but a two-body continuum state 
with $\bar{K}$ and $\bar{K}N$ quasibound state. 

Next we estimate the imaginary part of the energy $E$ for the total Hamiltonian $H$
by calculating the expectation value with the 
wave function $\Psi$ obtained by the Hamiltonian $H^{\rm Re}$:
\begin{equation}
E^{\rm Im}=\langle \Psi| {\rm Im}V_{\bar{K}N}|\Psi \rangle.
\end{equation}
The total decay width for $\bar{K}\bar{K}N$ 
is estimated as 
$\Gamma=-2 E^{\rm Im}$. In the present calculation, we have only three-body decays
to $\pi \Sigma \bar K$ and $\pi \Lambda \bar K$
by the model setting. 

We also calculate several quantities characterizing the structure of the three-body system,
such as spatial configurations of the constituent particles and probabilities to have specific
isospin configurations. These values are calculated as expectation values of the wave functions.
The root-mean-square (r.m.s.) radius of the $\bar K$ distribution is defined 
as the average of the distribution of each anti-kaon by
\begin{equation}
r_{\bar K}\equiv\sqrt{\left \langle \Psi\left|
{\textstyle \frac{1}{2}}({\bf x}^2_1+{\bf x}^2_2)\right|
\Psi \right\rangle},
\end{equation}
which is measured from the center of mass of the three-body system. 
For the two-body $\bar KN$ system, $r_{\bar K}$ is given by the spatial 
coordinate of $\bar K$, ${\bf x}_{\bar K}$, measured 
from the center of mass of the two-body system as  
$r_{\bar K}=\sqrt{\left \langle {\bf x}^2_{\bar K} \right\rangle}$.
We also
calculate the r.m.s.~value of the relative $\bar K$-$\bar K$ distance 
defined by
\begin{equation}
d_{\bar K\bar K}\equiv\sqrt{\left \langle \Psi\left|
{\bf r}_3^2\right|
\Psi \right\rangle}.
\end{equation}
The probabilities for the three-body system to have the isospin $I_{\bar K \bar K}$ states are introduced by
\begin{equation}
\Pi \left(\left[\bar K \bar K\right]_{I_{K K}}\right)
\equiv \left\langle \Psi \left|P_{\bar K\bar K}(I_{K K})\right|
\Psi \right\rangle, \label{eq:probKK}
\end{equation}
where $P_{\bar K\bar K}(I_{KK})$ is the projection operator
for the isospin configuration
$\left[[\bar{K}\bar{K}]_{I_{KK}}N\right]_{I=1/2}$, as given before.
We also calculate the r.m.s.\ radius of the $\bar K$ distribution
for the each isospin state
\begin{equation}
r_{\bar K}|_{I_{KK}}\equiv\sqrt{
\frac{
\left \langle \Psi\left|\frac{1}{2}({\bf x}^2_1+{\bf x}^2_2)\right|
P_{\bar K\bar K}(I_{KK})
\Psi \right\rangle}
{\left \langle \Psi\left|
P_{\bar K\bar K}(I_{KK})\right|
\Psi \right\rangle}
},
\end{equation}
which is normalized by Eq.~(\ref{eq:probKK}). 

In order to investigate the structure of the $\bar K \bar K N$ system
further,  
we calculate the expectation values with the unsymmetrized wave function $\Phi$ given in Eq.~(\ref{eq:wavefunc}). Although
these expectation values are not observable in the real $\bar K\bar K N$ system 
having two identical bosons, they are helpful to understand 
the structure of the three-body system and to investigate the symmetrization 
effect.
We calculate norms of the wave functions for the isospin
$\left[[\bar{K}_2N]_{I_{KN}}\bar K_1\right]_{I=1/2}$ states, where
the total isospin $1/2$ for the $\bar K \bar K N$ system
is given by combination of total isospin $I_{KN}$ for the $\bar K_2 N$
subsystem and isospin $1/2$ for the $\bar K_1$,
\begin{equation}
\Pi \left(\left[\bar K_2N\right]_{I_{KN}}\right)_{\Phi}
\equiv \left\langle \Phi \left|P_{\bar K_2 N}(I_{KN})\right|
\Phi \right\rangle,
\end{equation}
where $P_{\bar K_2 N}(I_{KN})$ is again the isospin projection operator. Note that this {\it cannot} be interpreted  as a probability for the $\left[[\bar{K}_2N]_{I_{KN}}\bar K_1\right]_{I=1/2}$ states in the physical $\bar K \bar K N$ system, since the norm is not normalized to be unity without the symmetrization of the anti-kaon wave functions.  
The r.m.s. values of the  $\bar K_2$-$N$ distance and  
the $\bar K_2N$-$\bar K_1$
distance for $\Phi$ are evaluated as
\begin{subequations} 
\label{eq:distance}
\begin{eqnarray}
d_{\bar K_2N} &\equiv& \sqrt{
\frac{
\left \langle \Phi \left|{\bf r}^2_1\right|
\Phi \right\rangle}
{\left \langle \Phi| \Phi \right\rangle}
} , \\
d_{(\bar K_2N){\rm -}\bar K_1} &\equiv& \sqrt{
\frac{
\left \langle \Phi \left|{\bf R}^2_1\right|
\Phi \right\rangle}
{\left \langle \Phi| \Phi \right\rangle}
},
\end{eqnarray}
\end{subequations}
respectively. 

The perturbative treatment performed above is justified qualitatively
in the case of $|\langle \Psi| {\rm Im}V|\Psi \rangle| \ll
|\langle \Psi| {\rm Re}V|\Psi \rangle|$. 
For the two-body system $\bar{K}N$, we find the perturbative treatment 
good, observing  that 
$|\langle {\rm Im}V_{\bar{K}N} \rangle|= 22$ MeV is much smaller than
$|\langle {\rm Re}V_{\bar{K}N} \rangle|\sim 100$ MeV in the HW-HNJH potential
case, for instance. This is responsible for
that we get the reasonable energy
$E=1423-22 i$ MeV for the $\bar KN$ with the perturbative treatment.
(In a full calculation with the $\bar{K}N$ effective interaction,
the scattering amplitude reproduces the $\Lambda(1405)$ resonance at
1420 MeV with 40 MeV width)
Also in the case of the $\bar{K}\bar{K}N$ system, 
it is found that the absolute values of the
perturbative energy 
$|\langle \Psi| {\rm Im}V_{\bar{K}N}|\Psi \rangle|= 20\sim 30 $ MeV 
is much smaller than the real potential energy
$|\langle \Psi| {\rm Re}V_{\bar{K}N}+
{\rm Re}V_{\bar{K}\bar{K}}|\Psi \rangle| = 100\sim 200$ MeV
in the present calculations. This is because the dominant 
contribution of $\langle V_{\bar K N}\rangle$ 
comes from $I=0$ channel which has
the strong attractive potential
with the weak imaginary part compared
with the real part.
 
\section{Results} \label{sec:results}

In this section, we show the results of 
calculation of the $\bar K \bar K N$ system with $I=1/2$ and 
$J^P=1/2^+$.
For the $\bar K N$ interactions, 
we compare two potential, the HW-HNJH potential
and the AY potential, as discussed in the previous sections. 
We first show the calculation without the  $\bar{K}\bar{K}$ 
interactions $V_{\bar{K}\bar{K}}$, in order to see if the $\bar KN$
interaction is strong enough for binding the three-body system.  
Later in subsection \ref{sec:VKK}, we discuss the effect of 
$V_{\bar{K}\bar{K}}$ by introducing possible repulsive 
$\bar{K}\bar{K}$ interactions in the $I=1$ channel.


\subsection{Bound $\bar{K}\bar{K}N$ state 
without $\bar{K}\bar{K}$ interaction}
In this subsection, we present the results of the 
$\bar{K}\bar{K}N$ state calculated 
without the $V_{\bar{K}\bar{K}}$ interaction. 

\subsubsection{Energy, width and decay modes of $\bar{K}\bar{K}N$ state}

\begin{figure}[th]
\centerline{\includegraphics[width=6.5cm]{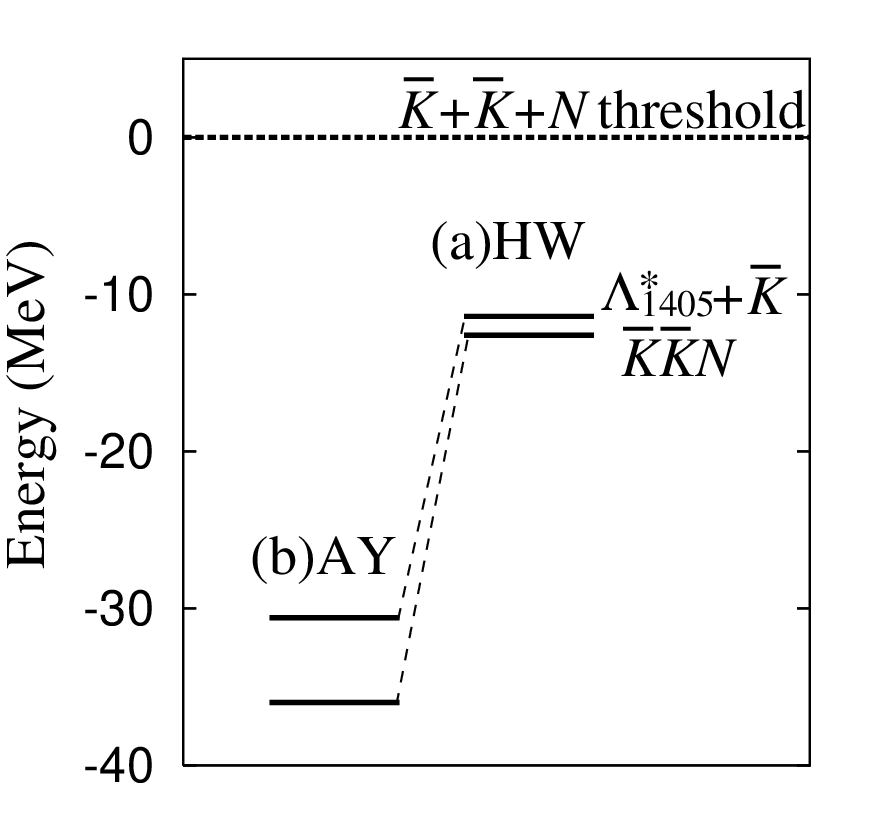}}
\caption{\label{fig:spe} Energies of the $\bar{K}\bar{K}N$
calculated with (a)~the HW-HNJH ($\delta\omega=11$ MeV) potential
and (b)~the AY potential without the
$\bar{K}\bar{K}$ interaction. The calculated thresholds of 
$\bar K$ and the quasibound $\bar{K}N$ state
are also shown. The $\bar K$+$\bar K$+$N$ threshold is located at 1930 MeV.}
\end{figure}

First of all, it is very interesting that 
the $\bar{K}\bar{K}N$ bound 
state is obtained below the threshold of the $\bar K$ and the quasibound $\bar KN$ state in both calculations with  the 
HW-HNJH and with the AY potentials, as seen in
Fig.~\ref{fig:spe}, where we show the level structure of the 
$\bar K \bar K N$ system.
In Table \ref{tab:energy}, we show the results for the energies and
radius of the $\bar K \bar K N$ state
as well as those for the $\bar{K}N$ state.
For the energy-dependent HW-HNJH potential, 
we take two energies $\delta\omega=0$ and 11 MeV
with $\omega=M_N+M_K-\delta\omega$.
In the table, $B(\bar K N)$ and $B(\bar K \bar K N)$ denote
the binding energies for the $\bar KN$ and $\bar K \bar KN$
systems measured by the energies of the two-body and 
three-body break-up states, respectively. 

The $\bar K \bar K N$ bound state appears as small as 1 MeV below the 
threshold  of the $\bar K$ and  quasibound  $\bar{K}N$ state ($-B(\bar{K}N)$)
in both cases of the HW-HNJH potential.
The calculated value for $r_{\bar K}$ is about 4 fm, which is much 
larger than that of the two-body $\bar{K}N$ bound state. 
It indicates that
the $\bar{K}\bar{K}N$ state is loosely bound with a large radius. 
In the case of the AY potential, the $\bar{K}\bar{K}N$ energy 
is $-36$ MeV, which is 5 MeV below the $\bar{K}N$+$\bar{K}$
threshold ($-31$ MeV).
The AY potential
gives a deeper binding and a
smaller radius, $r_{\bar K}=2$ fm than the HW-HNJH potential,
reflecting  the stronger $\bar{K}N$ $(I=0)$ attraction 
in the AY potential. 

The $\bar{K}\bar{K}N\rightarrow \pi Y\bar{K}$ decay width 
of the $\bar{K}\bar{K}N$ state 
is evaluated by the imaginary $\bar{K}N$ potentials 
as $\Gamma(\bar{K}\bar{K}N\rightarrow \pi Y\bar{K})=-2 E^{\rm Im}$. 
In the present results, 
we obtain the width in the range 
of $51$ MeV to $57$ MeV. This  implies that the 
$\bar{K}\bar{K}N$ state has a comparable width to 
that of the $\bar{K}N$ state for $\Lambda(1405)$.
It is interesting to note that 
the dominant contribution in the  decay width of the $\bar K \bar K N$ state 
comes from  the $I=0$ channel for the $\bar KN$ subsystem,
while the $I=1$ channel gives only a few MeV contribution.
This means that the $\bar K \bar K N$ state  has 
dominant $\pi\Sigma \bar{K}$ decay and
relatively small $\pi\Lambda \bar{K}$ decay modes.
This characteristic decay pattern comes from the strong attraction 
in the $\bar K N$ channel with $I=0$ reproducing the $\Lambda(1405)$
as a quasibound state. 

\subsubsection{Structure of $\bar{K}\bar{K}N$ state}

Let us discuss  the binding mechanism and the structure of the $\bar{K}\bar{K}N$ 
system with $I=1/2$ in detail. 
For this purpose, we  analyze the 
wave functions obtained in the present few-body calculation in terms of the
isospin configuration and the spatial structure of the  $\bar{K}\bar{K}N$
system, since the potential energy is determined by 
both of the isospin and spatial structure.

We first discuss the binding mechanism based on the isospin configurations. 
For convenience, we introduce the following notations; 
$[\bar{K}\bar{K}]_{I}$ denotes the isospin 
configuration $\left[[\bar{K}\bar{K}]_{I_{KK}=I}N\right]_{I=1/2}$
for the two-body $\bar K \bar K$ subsystem with $I=0$ or~$1$ 
in the three-body system,
and $[\bar{K}N]_0$ stands for $\left[[\bar{K}N]_{I_{KN}=0}\bar{K}\right]_{I=1/2}$
for the two-body $\bar K N$ subsystem with $I=0$, which is 
$\Lambda(1405)+\bar{K}$-like
isospin configuration.

According to the following argument, it is easy to understand that
the $[\bar{K}\bar{K}]_1$ and $[\bar{K}N]_0$ configurations are energetically 
favoured to gain potential energy of the $\bar{K}N$ subsystem. 
The effective $\bar{K}N$ potential $V^{I}_{\bar{K}N}$
has strong attraction in the $I=0$ and weak attraction
in the $I=1$ channel.
From group-theoretical arguments it is easily found that, on one hand, 
the former configuration $[\bar{K}\bar{K}]_1$ consists of 
$[\bar{K}N]_0$ and $[\bar{K}N]_1$ components with a ratio 3:1.
As a result, 
both of two anti-kaons effectively feel the potential 
$\frac{3}{4}V^{I=0}_{\bar{K}N}+\frac{1}{4}V^{I=1}_{\bar{K}N}$, which 
is moderate attraction. 
On the other hand, in the latter configuration $[\bar{K}N]_0$, 
one of the anti-kaons couples with the nucleon in $I_{KN}=0$ and 
gains the strong attraction of $V^{I=0}_{\bar{K}N}$, and, at the same time,
the other anti-kaon feels much weaker attraction as 
$\frac{1}{4}V^{I=0}_{\bar{K}N}+\frac{3}{4}V^{I=1}_{\bar{K}N}$.

In the present calculation, 
it is found that the probability $\Pi([KK]_1)$ for the three-body system to have the $[\bar{K}\bar{K}]_1$ configuration is
dominant in the $\bar{K}\bar{K}N$ wave functions 
as shown in Table \ref{tab:norm}, in which 
$\Pi([\bar K\bar K]_1)$ is 0.87 for the result of the 
HW-HNJH~($\delta\omega=11$ MeV) potential, and it is 
0.91 in the case of the AY potential. 
These values are in between two limits, $[\bar{K}N]_0$ and $[\bar K \bar K]_1$;
In the $[\bar{K}N]_0$ limit,
the probability $\Pi([\bar K\bar K]_1)$ should be 0.75
when the symmetrization of two anti-kaons is ignored, 
while $\Pi([\bar K \bar K]_1)=1$ for the pure $[\bar{K}\bar{K}]_1$ state.
The present calculation implies that 
the $\bar{K}\bar{K}N$ state is regarded as an admixture of the 
isospin configurations 
$[\bar{K}\bar{K}]_1$ and $[\bar{K}N]_0$.
Therefore,  the rearrangement of the
isospin configurations  
is essential in the $\bar{K}\bar{K}N$ bound state.

Next we discuss the spatial structure of 
the $\bar{K}\bar{K}N$ state.
Since two anti-kaons are identical bosonic particles, 
$\bar K_1$ and $\bar K_2$
cannot be identified in the symmetrized wave function $\Psi$.
For an intuitive understanding,  
it is helpful to analyze  the
wave function $\Phi$ obtained before the symmetrization.
For the wave function $\Phi$, 
we can definitely calculate the expectation values for 
the $\bar{K}_2$-$N$ and $(\bar K_2N)$-$\bar K_1$ distances
in the rearrangement channel $c=1$ as given by Eq.~(\ref{eq:distance}).
In the calculated results, it is found that $d_{\bar{K}_2N}$
is almost the same as the $\bar{K}$-$N$ distance in the 
$\bar{K}N$ quasibound state, while 
$d_{(\bar K_2N){\rm -}\bar K_1}$ is 
remarkably large. Actually, $d_{(\bar K_2N){\rm -}\bar K_1}$
is more than three times larger than $d_{\bar{K}_2N}$
in all choices of the interactions as shown in Table.~\ref{tab:norm}.
For example, we obtain $d_{\bar{K}_2N}=1.6$ fm 
and $d_{(\bar K_2N){\rm -}\bar K_1}=6.2$ fm 
with the HW-HNJH ($\delta\omega$=11 MeV) 
potential.
It indicates that one of the kaons widely distributes around the 
nucleon with very loosely binding
and the other kaon is moving in the vicinity of the nucleon.
In addition,  the wave function $\Phi$  
contains  dominantly the  $[\bar{K}_2N]_0$ component
and little the $[\bar{K}_2N]_1$ configuration,
which are shown as the norms of the corresponding wave functions in Table\ref{tab:norm}.
This means that the $\bar K_2 N$ subsystem has more likely the $I=0$ 
component and, thus the $\Phi$ has a $\Lambda(1405)+\bar{K_{1}}$ cluster
structure \footnote{The idea of 
$\Lambda(1405)$ cluster in kaonic nuclei was proposed in the 
$K^-pp$ system in Ref.~\cite{Yamazaki:2007hj}.}. 
It is worth noting that,
even though the group theoretical argument suggests that the probability $\Pi([\bar K \bar K]_1)$ for the normalized wave function  should be 0.75 without the symmetrization,
the $[\bar{K}\bar{K}]_1$ component is obtained as $\Pi([KK]_1)\sim 0.9$ 
for the symmetrized  wave function $\Psi$, as discussed above. This is because of the symmetrization effect. 

\begin{figure}[th]
\centerline{\includegraphics[width=7.5cm]{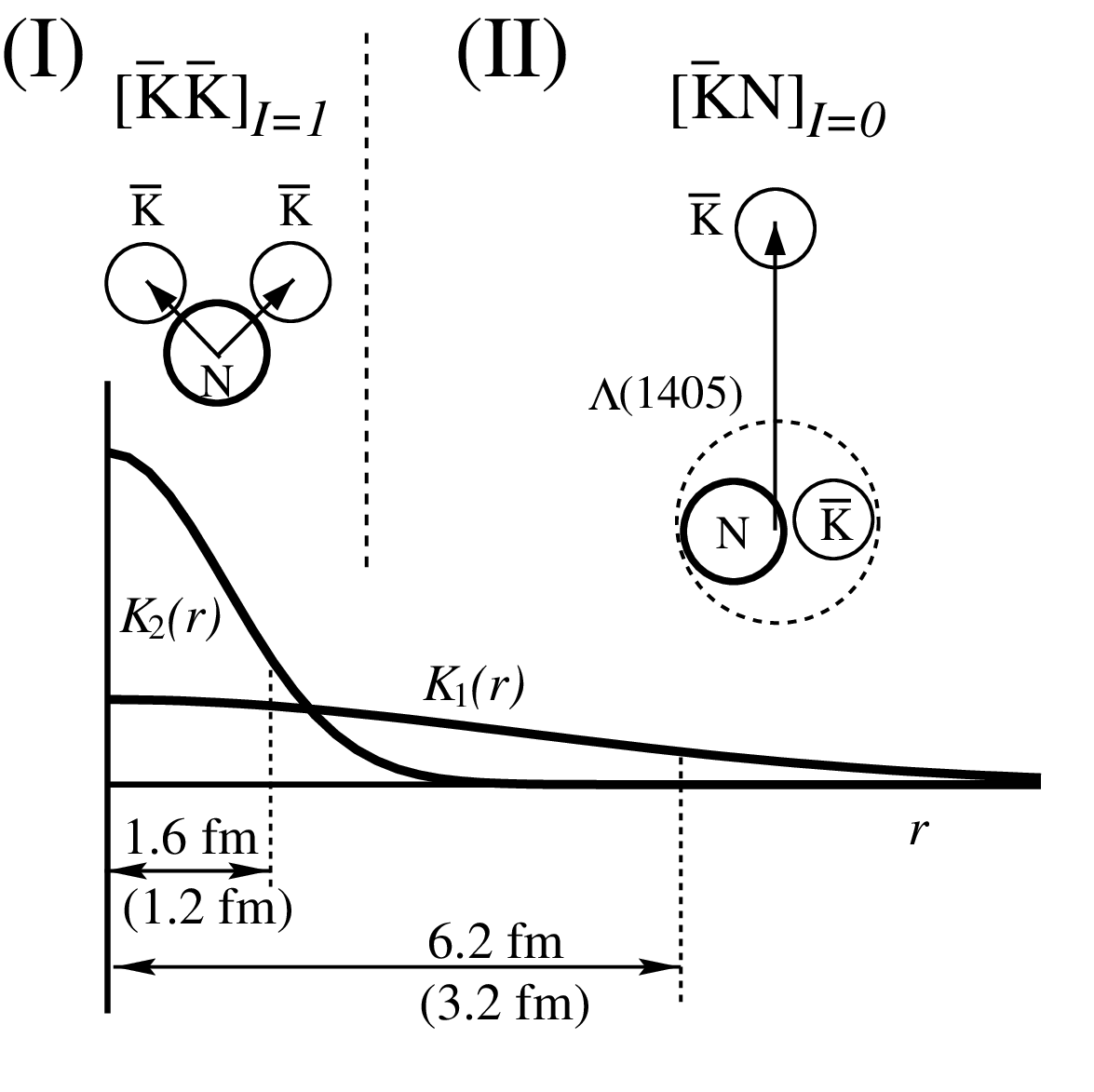}}
\caption{\label{fig:kkp} 
Schematic figure for the $\bar{K}\bar{K}N$ state. 
$K_1(r)$ and $K_2(r)$ indicate the anti-kaon wave functions.
One of the anti-kaons distributes in a wide region 
and the other anti-kaon distributes near the nucleon.
The values shown in the figure without the parenthesis
are $d_{\bar{K}_2N}$ and $d_{(\bar K_2N){\rm -}\bar K_1}$
calculated with the HW-HNJH potential ($\delta\omega=11$ MeV), while
the values in the parentheses are those for the AY
potential. 
In the inner region (I),
two anti-kaons are coupled to 
isospin symmetric with the isospin configuration
$[[\bar{K}\bar{K}]_{I_{KK}=1}N]_{I=1/2}$ because 
they occupy the same orbit and are spatial symmetric.
In the outer region (II), 
the nucleon and one of the anti-kaons form a $\Lambda(1405)$ state and
the other anti-kaon is moving around the $\Lambda(1405)$.}
\end{figure}

Now let us combine the analysis of the isospin and spatial configurations.
The $\bar{K}\bar{K}N$ bound state can be 
interpreted as a hybrid of the two configurations:
the $[\bar{K}\bar{K}]_1$ in the inner region (I) 
and the $\Lambda(1405)+\bar{K}$ cluster 
in the asymptotic region (II) as shown in the
schematic figure (Fig.~\ref{fig:kkp}).
As mentioned above, 
one of the anti-kaons distributes in a spatially wide region 
and the other anti-kaon distributes near the nucleon.
In the inner region~(I),
two anti-kaons are coupled to 
isospin symmetric with the isospin configuration
$[\bar{K}\bar{K}]_1$ because 
they occupy the same orbit and are spatially symmetric.
In the outer region~(II), 
the nucleon and one of the anti-kaons form a $\Lambda(1405)$ state and
the other anti-kaon is moving around the $\Lambda(1405)$.
This kaon-halo like structure is reminiscent of the neutron-halo observed 
in unstable nuclei.

\begin{table*}[ht]
\caption{ 
\protect\label{tab:energy} 
Energies and root-mean-square~(r.m.s.) radii and distances of the 
$\bar{K}N$ and the $\bar{K}\bar{K}N$ states
calculated without 
the $\bar{K}\bar{K}$ interaction. For the $\bar{K}N$ interactions,
the AY potential and the HW-HNJH potential with the energy parameter
$\omega=M_N+M_K-\delta\omega$ (MeV) are used. 
For the $\bar{K}N$ state,
the real energy $E^{\rm }=-B(\bar{K}N)$, the imaginary energy $E^{\rm Im}$,
the r.m.s.~$\bar{K}$-$N$ distances($d_{\bar KN}$),
$r_{\bar K}$ are shown.
For the $\bar{K}\bar{K}N$ state,
the real energy $E^{\rm }=-B(\bar{K}\bar{K}N)$, the energy 
relative to the $\bar{K}N+\bar{K}$ threshold,
the imaginary energy $E^{\rm Im}$, 
$r_{\bar K}$ are shown.
We also list the expectation values of ${\rm Im} V^{I=0}_{\bar{K}N}$ and 
${\rm Im} V^{I=1}_{\bar{K}N}$ separately.}
\begin{tabular}{lrrr}
\hline
& \qquad AY  &  \multicolumn{2}{c}{HW-HNJH} \\
&  & \qquad$\delta\omega=0$ & \qquad$\delta\omega=11$ \\
\hline
\multicolumn{3}{c}{$\bar{K}N(I=0)$ state}\\ 
\hline
$-B(\bar{K}N)$ (MeV) 
&	$-$30.6 	&	$-$10.4 	&	$-$11.4 \\
$E^{\rm Im}$  (MeV) 
&	$-$19.9 	&	$-$20.9 	&	$-$21.8 \\
$d_{\bar KN}$ (fm) 
&	1.4 	&	2.0 	&	1.9 	\\
$r_{\bar K}$ (fm) 
&	0.9 	&	1.3 	&	1.2 	\\
\hline
\multicolumn{3}{c}{$\bar{K}\bar{K}N(I=1/2)$ state}\\ 
\hline
$-B(\bar{K}\bar{K}N)$ (MeV) 
&	$-$36.0 	&	$-$11.3 	&	$-$12.6 	\\
$-B(\bar{K}\bar{K}N)+B(\bar{K}N)$ (MeV) 
&	$-$5.4 	&	$-$1.0 	&	$-$1.2 	\\
$E^{\rm Im}$ (MeV) 
&	$-$28.3 	&	$-$25.4 	&	$-$26.8 	\\
$\langle {\rm Im}V^{I=0} \rangle$ (MeV) 
&	$-$24.8 	&	$-$24.2 	&	$-$25.5 	\\
$\langle {\rm Im}V^{I=1} \rangle$ (MeV) 
&	$-$3.5 	&	$-$1.2 	&	$-$1.3 	\\
$r_{\bar K}$ (fm) 
&	2.0 	&	4.2 	&	3.8 	\\
\hline
\end{tabular}
\end{table*}

\begin{table*}[ht]
\caption{ 
\protect\label{tab:norm} 
Properties such as the isospin configurations and radii
in the three-body $\bar K \bar K N$ system
calculated without the $\bar{K}\bar{K}$ interaction.
In the upper part, we show the values in the total wave function $\Psi$ 
obtained after the symmetrization. 
The probabilities $P([\bar K \bar K]_{0,1})$ to have 
the $[\bar K \bar K]_{0,1}$ configurations are listed.
The $r_{\bar K}$ in the 
$[\bar{K}\bar{K}]_0$ and 
$[\bar{K}\bar{K}]_1$ components are shown separately. 
The calculated values for $d_{\bar K\bar K}$ are also shown.
We also present the expectation values for the unsymmetrized 
wave function $\Phi$.
The norms of the wave functions for the isospin configurations  
$[\bar{K}_2N]_0$ and 
$[\bar{K}_2N]_1$, and
distances $d_{\bar K_2N}$ and $d_{(\bar K_2N){\rm -}\bar K_1}$
calculated with the wave function $\Phi$ are shown.
}
\begin{tabular}{lrrr}
\hline
& \qquad AY  &  \multicolumn{2}{c}{HW-HNJH} \\
&  & \qquad$\delta\omega=0$ & \qquad$\delta\omega=11$\\
\hline
\multicolumn{3}{c}{expectation values for $\Psi$}\\ 
\hline
$\Pi([\bar K\bar K]_0)$
&	0.09 	&	0.13 	&	0.13 	\\
$\Pi([\bar K\bar K]_1)$
&	0.91 	&	0.87 	&	0.87 	\\
$r_{\bar K}|_{I_{KK}=0}$ (fm)
&	2.7 	&	5.2 	&	4.8 	\\
$r_{\bar K}|_{I_{KK}=1}$ (fm)
&	1.9 	&	4.0 	&	3.7 	\\
$d_{\bar K\bar K}$ (fm)
&	3.1 	&	6.4 	&	5.9 	\\
\hline
\multicolumn{3}{c}{expectation values for $\Phi$}\\ 
\hline
$\Pi([\bar K_{2}N]_{0})_{\Phi}$
&	0.80 	&	0.81 	&	0.80 	\\
$\Pi([\bar K_{2}N]_{1})_{\Phi}$
&	0.00 	&	0.00 	&	0.00 	\\
$d_{\bar K_2N}$ (fm)
&	1.2 	&	1.7 	&	1.6 	\\
$d_{(\bar K_2N){\rm -}\bar K_1}$ (fm)
&	3.2 	&	6.8 	&	6.2 	\\
\hline
\end{tabular}
\end{table*}

\subsection{effect of repulsive $\bar{K}\bar{K}$ interaction}
\label{sec:VKK}
In the previous subsection, we presented the results calculated 
without $\bar{K}\bar{K}$ interactions.
We discuss possible effects of the $\bar{K}\bar{K}$ interactions.
As already mentioned, the $I=0$ is forbidden for 
$s$-wave $\bar{K}\bar{K}$ states due to Einstein-Bose statistics.
We here investigate how the repulsive $\bar{K}\bar{K}$ interactions 
may affect the $\bar{K}\bar{K}N$ state
by introducing $V^{I}_{\bar{K}\bar{K}}$ for the $I=1$ channel.

We use the $V^{I=1}_{\bar{K}\bar{K}}$ interactions adjusted to 
reproduce the scattering length $a^{I=1}_{KK}=-0.14$ fm  of lattice
QCD calculation. We also use a weaker repulsion with $a^{I=1}_{KK}=-0.10$ fm.
The calculated results with 
$V^{I=1}_{\bar{K}\bar{K}}$ are shown in Table \ref{tab:VKK}.
It is found that the $\bar{K}\bar{K}N$ bound state is obtained
even with the possible repulsive $\bar{K}\bar{K}$ interactions. 
Because of the repulsive $\bar{K}\bar{K}$ interactions, 
the $\Lambda(1405)+\bar K$ cluster develops more and
the anti-kaon is further loosely bound.
As a result, the absolute value of the imaginary energy ${\rm Im}E$ decreases
as shown in Table \ref{tab:VKK}.
The value of $d_{(\bar K_2N){\rm -}\bar K_1}$
is extremely large as $d_{(\bar K_2N){\rm -}\bar K_1} \ge 16$ fm in
the HW-HNJH potential, and the 
$\Lambda(1405)+\bar K$ cluster feature becomes more remarkable.

There still remains ambiguity of the strengths of the 
$\bar{K}\bar{K}$ interactions due to few experimental 
data. Here we estimate possible boundaries of the repulsive interaction 
for the formation of the bound $\bar K \bar K N$ state.  
For the AY potential, 
the $\bar{K}\bar{K}N$ state is still bound 
even with a further strong $\bar{K}\bar{K}$ interaction 
as $a^{I=1}_{KK}=-0.20$ fm. In 
the case of the HW-HNJH potential, 
the $\bar{K}\bar{K}N$ energy is found above the $\bar{K}N+\bar{K}$
threshold  with more repulsive $\bar{K}\bar{K}$ interactions than $a^{I=1}_{KK}=-0.15$ fm. 
In such a case, a resonance state 
or a virtual state may appear near the threshold.
This might have a larger width than that of 
the three-body bound state.
For the detailed structure of such states, we need to use formulations 
beyond the present framework, 
because the continuum states
are not taken into account in the present calculations.

\begin{table*}[ht]
\caption{ 
\protect\label{tab:VKK} 
Theoretical results of the $\bar{K}\bar{K}N$ state
calculated with the $\bar{K}\bar{K}$ interactions.
The range parameters of the $\bar{K}\bar{K}$ interactions are taken to be
$b=0.47$ fm and $b=0.66$ fm for the HW-HNJH and the AY potentials, 
respectively.
Energies and root-mean-square (r.m.s.) radii and distances 
are listed.
The expectation values for the various isospin configurations, and
those for the unsymmetrized wave function $\Phi$ are also
shown. The detailed descriptions are given in captions of 
Tables \ref{tab:energy} and \ref{tab:norm}.
In the calculations for HW-HNJH($\delta\omega=0$ MeV) with
$U^{I=1}_{KK}=70$ MeV, we take the extended basis space as
$(n_{\rm max},N_{\rm max})=(15,28)$ and
$(r_{\rm min},r_{\rm max},R_{\rm min},R_{\rm max})=(0.2,20,0.2,400)$
(fm) for the configuration $c=1$ to get the convergent solution.
}
\begin{tabular}{lrrrrrr}
\hline
& \multicolumn{2}{c}{AY+ $V_{\bar{K}\bar{K}}$} &  \multicolumn{4}{c}{HW-HNJH 
+ $V_{\bar{K}\bar{K}}$} \\
& &  & \multicolumn{2}{c}{$\delta\omega=0$} & 
\multicolumn{2}{c}{$\delta\omega=11$} \\
$a^{I=1}_{KK}$ (fm) & 1.0 & 1.4 & 1.0 & 1.4 & 1.0 & 1.4 \\
$U^{I=1}_{KK}$ (MeV) & 70 & 104 & 205 & 313 & 205 & 313\\
\hline
\multicolumn{6}{c}{$\bar{K}\bar{K}N(I=1/2)$ state}\\ 
\hline
$-B(\bar{K}\bar{K}N)$ (MeV) 
&	$-$33.2 	&	$-$32.3 	&	$-$10.5 	&	$-$10.4 	&	$-$11.5 	&	$-$11.4 	\\
$-B(\bar{K}\bar{K}N)+B(\bar{K}N)$ (MeV) 
&	$-$2.6 	&	$-$1.7 	&	$-$0.1 	&	$-$0.005 	&	$-$0.1 	&	$-$0.01 	\\
$E^{\rm Im}$ (MeV) 
&	$-$25.1 	&	$-$23.8 	&	$-$21.7 	&	$-$21.0 	&	$-$22.8 	&	$-$22.0 	\\
$\langle {\rm Im}V^{I=0} \rangle$ (MeV) 
&	$-$22.9 	&	$-$22.1 	&	$-$21.5 	&	$-$21.0 	&	$-$22.5 	&	$-$22.0 	\\
$\langle {\rm Im}V^{I=1} \rangle$ (MeV)
&	$-$2.2 	&	$-$1.7 	&	$-$0.2 	&	0.0 	&	$-$0.3 	&	$-$0.1 	\\
$r_{\bar{K}}$ (fm) 
&	2.7 	&	3.2 	&	12 	&	49 	&	10 	&	32 	\\
\hline
\multicolumn{6}{c}{expectation values for $\Psi$}\\ 
\hline
$\Pi([\bar K \bar K ]_{0})$
&	0.13 	&	0.14 	&	0.20 	&	0.24 	&	0.20 	&	0.23 	\\
$\Pi([\bar K \bar K ]_{1})$
&	0.87 	&	0.86 	&	0.80 	&	0.76 	&	0.80 	&	0.77 	\\
$r_{\bar K}|_{I_{KK}=0}$ (fm)
&	3.4 	&	3.9 	&	13 	&	50 	&	11 	&	33 	\\
$r_{\bar K}|_{I_{KK}=1}$ (fm)
&	2.6 	&	3.0 	&	12 	&	48 	&	9.8 	&	31 	\\
$d_{\bar K\bar K}$ (fm)
&	4.1 	&	4.9 	&	18 	&	76 	&	16 	&	49 	\\
\hline
\multicolumn{4}{c}{expectation values for $\Phi$}\\ 
\hline
$\Pi([\bar K_{2}N]_{0})_{\Phi}$
&	0.85 	&	0.87 	&	0.93 	&	0.99 	&	0.92 	&	0.98 	\\
$\Pi([\bar K_{2}N]_{1})_{\Phi}$
&	0.00 	&	0.00 	&	0.00 	&	0.00 	&	0.00 	&	0.00 	\\
$d_{\bar K_2N}$ (fm)
&	1.2 	&	1.3 	&	1.9 	&	2.0 	&	1.8 	&	1.9 	\\
$d_{(\bar K_2N){\rm -}\bar K_1}$ (fm)
&	4.2 	&	5.0 	&	19 	&	76 	&	16 	&	49 	\\
\hline
\end{tabular}
\end{table*}

\section{Summary and concluding remarks} \label{sec:summary}
We have investigated the $\bar{K}\bar{K}N$ system with $I=1/2$ and $J^P=1/2^+$ 
as an example of multi-hadron systems with anti-kaons. 
We have performed  a non-relativistic three-body calculation 
by using the effective $\bar{K}N$ interactions
proposed by Hyodo-Weise and Akaishi-Yamazaki.
With the $\bar{K}N$ potentials and no $\bar{K}\bar{K}$ interaction, 
the present calculation suggests that 
a weakly bound $\bar{K}\bar{K}N$ state can be formed 
below the $\bar{K}N$+$\bar{K}$ threshold energy.
The AY potential for the $\bar{K}N$ interactions provides  
a deeper bound state of the $\bar{K}\bar{K}N$ system 
because of the stronger attraction in the  $\bar{K}N$ channel with $I=0$
than the case of the HW potential.

Investigating the wave function obtained by the three-body calculation, 
we have found that 
the $\bar{K}\bar{K}N$ bound state can be 
interpreted as a hybrid of two configurations;
In the inner region, two kaons are spatially symmetric and 
couples to $I_{KK}=1$, while 
in the asymptotic region where one kaon far from the nucleon, 
the system is regarded as the $\Lambda(1405)$+$\bar K$ like cluster.
The root-mean-square radius of $\bar{K}$ distribution 
is found to be a large
value due to the $\Lambda(1405)$+$\bar K$ like component 
with a loosely bound kaon around the $\Lambda(1405)$.
This kaon-halo like structure is reminiscent of 
the neutron-halo observed in unstable nuclei. 

We have also evaluated the decay width of the $\bar K \bar K N$ system
to $\pi Y\bar K$, obtaining $\Gamma= 40\sim60$ MeV, which is comparable 
to the width of $\Lambda(1405)$. The dominant mode is found to be 
$\bar{K}\bar{K}N\rightarrow \bar K (\pi \Sigma)_{I=0}$, reflecting the 
$\bar K$+$\Lambda(1405)$ cluster structure.
In this estimation, 
we have not considered three-body forces nor transitions to two-hadron decays. 
Nevertheless, such effects are expected to be suppressed 
because 
the overlap of the wave functions of three particles 
in a compact region is very small in the present case that
one of the $\bar{K}$s is loosely bound and 
distributes very widely around the $\bar{K}N$ subsystem.  

To estimate the effect of unknown $\bar{K}\bar{K}$ interactions,
we have introduced a repulsive interaction with $I=1$ which reproduces
the $K^+K^+$ scattering length obtained by lattice QCD calculation.
It is interesting that a bound state of the $\bar{K}\bar{K}N$ system is
possible even with the repulsive $\bar{K}\bar{K}$ interactions. 
The repulsive nature of the $\bar{K}\bar{K}$ interaction suggests that 
the $\Lambda(1405)$+$\bar K$ cluster develops more and
the anti-kaon is further loosely bound.

The peculiar structure that the wave function of one of the anti-kaon
spreads for long distance is a consequence that
the $\bar K N$ interaction is
strong enough to form a quasibound $\bar K N$ and $\bar K\bar K N$ states,
but is not so strong for the deeply bound $\bar K\bar K N$ state. 
This is very significant
for the multi-kaon system. In such systems, some of the anti-kaon could be
bound very loosely, so that many-body absorptions of anti-kaons
are suppressed and the anti-kaons keep their identity.

So far the experimental data for the $S=-2$ channel have been very poor.
In near future, detailed experimental investigations could clarify 
the structure of excited $\Xi$ baryons, for instance, in double strangeness 
reactions at J-PARC. 
The $\bar K \bar K N$ molecular state suggested in the present calculation 
is one of the excited $\Xi$ baryon with $J^{P}=1/2^{+}$ sitting around 1.9~GeV
and having characteristic properties. The main decay mode
is the three-body decay of the $\bar K \pi \Sigma$ with $I=0$ for the final 
$\pi \Sigma$, since the $\Lambda(1405)$ component in the three-body system
is a doorway for the decay. In addition,  productions of such a multi-hadron
system may be strongly dependent on transfered momentum, since the 
loosely bound state has a large spatial distribution, which leads a softer form factor. 
These would be good indications for identifying the 
molecular state in experiments.

In the present work, we have discussed the bound $\bar{K}\bar{K}N$ 
state obtained by the the single-channel 
three-body calculation, where 
effects of coupled-channel 
meson-baryon interactions are taken into account as 
effective single-channel $\bar{K}N$ interactions. 
Such coupled-channel effects could debase clear resonance shape 
for the three-body quasi-bound state in the spectrum, 
or could push the quasi-bound
state up to the two-body continuum and the 
quasi-bound state could be a virtual state. 
Nevertheless, in principle, the spectra which will 
be observed in experiments have information of resonances 
and virtual states.
For complete understanding of the $\bar K \bar K N$ molecule 
state, further detailed studies involving dynamics 
of three-body resonances
are necessary.
For instance, coupled-channel calculations of the three-body 
system including 
$\bar{K}N\leftrightarrow \pi\Sigma$~\cite{ikeda07-jps}, 
such as Faddeev type 
calculations done for the $ppK^{-}$ system in Refs.~\cite{shevchenko07,ikeda07}
and for two-meson and one baryon system with $S=-1$ in Refs.~\cite{MartinezTorres:2007sr},  will be required to obtain
qualitatively precise values of the energy and width.


Our suggestion of the $\bar{K}\bar{K}N$ quasibound state 
implies a possible existence of an excited baryon
with a molecule structure of two mesons surrounding
a baryon.
The present investigation may be an important step
to lead fundamental information 
on the physics of multi strange systems, such as
anti-kaons  in nuclear medium.


%


\section*{Acknowledgments}

The authors would like to thank Dr.~Hyodo for valuable discussions. 
They are also thankful to members of 
Yukawa Institute for Theoretical Physics (YITP)
and Department of Physics in Kyoto University, especially for
fruitful discussions.
%
This work is supported in part by
the Grant for Scientific Research (No.~18540263 and
No.~20028004) from Japan Society for the Promotion of Science (JSPS)
and from the Ministry of Education, Culture,
Sports, Science and Technology (MEXT) of Japan
%
A part of this work is done in the Yukawa International Project for 
Quark-Hadron Sciences (YIPQS).
The computational calculations of the present work were done by
using the supercomputer at YITP.

\end{document}